\documentclass[10pt]{article}
\usepackage{amsfonts}
\usepackage{mathrsfs}
\usepackage{amsmath}	
\usepackage{cite}
\usepackage{graphicx}
\usepackage{rotating}
\usepackage{hyperref}
\usepackage{verbatim}
\usepackage[all]{xy}

\csname @addtoreset\endcsname{equation}{section}
\newcommand{\beq}{\begin{equation}}
\newcommand{\eeq}{\end{equation}}
\newcommand{\bea}{\begin{eqnarray}}
\newcommand{\eea}{\end{eqnarray}}
\newcommand{\ba}{\begin{array}}
\newcommand{\ea}{\end{array}}
\newcommand{\bit}{\begin{itemize}}
\newcommand{\eit}{\end{itemize}}
\newcommand{\nn}{\nonumber}

\newcommand{\mezzo}{\frac{1}{2}}
\newcommand{\complesso}{{\ \hbox{{\rm I}\kern-.6em\hbox{\bf C}}}}
\newcommand{\reale}{{\hbox{{\rm I}\kern-.2em\hbox{\rm R}}}}
\newcommand{\uno}{ \,  \raisebox{+0.14em}{{\hbox{{\rm \scriptsize ]}} \raisebox{-0.2em}{\kern-.8em\hbox{1}}}} \, }  

\newcommand{\p}{\partial}

\renewcommand{\a}{\alpha}
\renewcommand{\b}{\beta}
\newcommand{\g}{\gamma}

\newcommand{\G}{\Gamma}
\renewcommand{\d}{\delta}
\newcommand{\D}{\Delta}
\newcommand{\e}{\epsilon}

\renewcommand{\k}{\kappa}
\renewcommand{\l}{\lambda}
\renewcommand{\L}{\Lambda}

\newcommand{\m}{\mu}

\newcommand{\n}{\nu}
\renewcommand{\r}{\rho}
\newcommand{\s}{\sigma}

\newcommand{\z}{\zeta}

\newcommand{\om}{\omega}
\newcommand{\Om}{\Omega}

\begin{document}


\begin{titlepage}
\begin{flushright}
UAI-PHY-16/06
\end{flushright}
\vspace{2.5cm}
\begin{center}
\renewcommand{\thefootnote}{\fnsymbol{footnote}}
{\huge \bf CFT Duals for Accelerating Black Holes}
\vskip 25mm
{\large {Marco Astorino\footnote{marco.astorino@gmail.com}}}\\
\renewcommand{\thefootnote}{\arabic{footnote}}
\setcounter{footnote}{0}
\vskip 10mm
{\small \textit{Departamento de Ciencias, Facultad de Artes Liberales,\\
Univesidad Adolfo Iba\~{n}ez, \\ 
Av. Padre Hurtado 750, Vi\~{n}a del Mar, Chile}
}
\end{center}
\vspace{5 cm}
\begin{center}
{\bf Abstract}
\end{center}
{The near horizon geometry of the rotating C-metric, describing accelerating Kerr-Newman black holes, is analysed. It is shown that, at extremality, even though not it is  isomorphic to the extremal Kerr-Newman, it remains a warped and twisted product of $AdS_2 \times S^2$. Therefore  the methods of the Kerr/CFT correspondence can successfully be applied to build a CFT dual model, whose entropy reproduce, through the Cardy formula, the Bekenstein-Hawking entropy of the accelerating black hole.\\
The mass of accelerating Kerr-Newman black hole, which fulfil the first law of thermodynamics, is presented.\\
Further generalisation in presence of an external Melvin-like magnetic field, used to regularise the conical singularity characteristic of the C-metrics, shows that the Kerr/CFT correspondence can be applied also for the accelerating and magnetised extremal black holes.
}
\end{titlepage}





\section{Introduction}
\label{intro}

In the last years there have been a great development of near horizon techniques to study the black hole physics \cite{horowitz}. These methods are being useful in the description of both macroscopic and microscopic properties of black holes in general relativity, especially at extremality. For instance the near horizon analysis was fundamental in the context of the Kerr/CFT correspondence \cite{strom08}, \cite{stro-duals}, \cite{strom-review} and \cite{compere-review}. While from a more classical point of view, the near horizon limit revealed also useful in determining the energy of magnetised black holes \cite{mass-mkn} and, through force-free electrodynamics, in modelling the Kerr black hole magnetosphere \cite{stro-magnsf} - \cite{oliv}, its accretion disk and jet dynamics, or describing some radiative processes around extreme Kerr black holes \cite{near-rad}, just to cite few relevant applications.\\
Here we will be mainly interested in the Kerr/CFT correspondence. It is based on the symmetries that emerge in the near horizon geometry, which usually are encoded in the $U(1) \times SL(2,\mathbb{R})$ group. Thanks to these symmetries it is possible to build a two dimensional conformal model dual to the gravitational one. From the features of the 2D CFT picture, some microscopical details of the black hole entropy can be extrapolated. In particular, through the Cardy formula it is possible to take into account the black hole microstates that generate their entropy.\\
Recently some generalisation of the Kerr/CFT correspondence have been discovered also for extremal black holes embedded in an external magnetic field, such as the Reissner-Nordstrom and Kerr(-Newman) spacetimes immersed in the Melvin magnetic universe \cite{micro-RN}-\cite{magn-kerr-cft}. In that case the near horizon geometry at extremality remains the same of the Kerr-Newman black hole. \\
The scope of this article is to further extend the applicability of the Kerr/CFT methods and to study possible generalisations of the Kerr-Newman near horizon geometry in case of extremal accelerating black holes. In this context the extremality plays a fundamental role because, at that specific parametric point, the event horizon symmetries are enhanced. This will be analysed in section \ref{acc-NH} and \ref{micro}. In particular we will focus on stationary and axisymmetric spacetimes. We will consider a subclass of the Demianski-Plebanski metrics \cite{plebanski}- \cite{griffiths}, known as C-metric and their rotating generalisation, often called rotating C-metric \cite{teo}\footnote{Rotating C-metrics admit also NUT charge, but in this work we will not consider it.}. These metrics are suitable to generalise the Kerr/CFT correspondence because they contain the (A)dS-Kerr-Newman spacetime, as a sub-case. In fact the rotating C-metric represents an (A)dS-Kerr-Newman black hole accelerating by the pressure of a pulling string (or pushing strut) \cite{podolsky}. Some basic properties of these metrics will be examined in section \ref{acc-rev}. In subsection \ref{mass-acc} we address a long standing open problem, that is the possibility of having a value for the mass of this accelerating Kerr-Newman spacetime compatible with standard laws black hole thermodynamics.  \\
Encouraged by the separability of the massless Klein-Gordon equation for probe scalar fields  on these accelerating black hole backgrounds, some speculations about the possibility of extend the correspondence with the conformal model also outside the extremal limit are presented in section \ref{micro}. \\
Since the string, that provide the acceleration, is mathematically represented by a conical singularity, in section \ref{magn-acc} we will confirm the validity of the above results by regularising the nodal singularity of the C-metric. The regularisation can be achieved, in the realm of the same Einstein-Maxwell theory, introducing an external background magnetic field that drives the black hole acceleration, in spite of the singular string. These kind of regularised metrics have been studied in the literature mainly in the context of pair creation of a black hole couple at expense of the external field energy \cite{gibbons-pair}, 
\cite{strominger-pair}, \cite{hawking-pair}  and \cite{rot-pair}.   \\



\section{Accelerating Kerr-Newman Black Hole Review}
\label{acc-rev}

Consider the action for  Einstein general relativity (without cosmological constant) coupled with standard Maxwell electromagnetism
\beq \label{EM-action}
             S[g_{\m\n}, A_\m]  =  - \frac{1}{16\pi}  \int_\mathcal{M} d^4x  \sqrt{-g} \left( R -  F_{\m\n} F^{\m\n} \right)    \ \  .
\eeq
Extremising it with respect to the metric and electromagnetic potential we get the following equations of motion for the metric and the gauge potential
\bea  \label{field-eq1}
                        &&   R_{\m\n} -   \frac{R}{2}  g_{\m\n} =  2  \left( F_{\m\r}F_\n^{\ \r} - \frac{1}{4} g_{\m\n} F_{\r\s} F^{\r\s} \right)  \quad ,       \\
           \label{field-eq2}             &&   \partial_\m \left( \sqrt{-g} F^{\m\n} \right) = 0   \quad ,
\eea
where, as usual, the Faraday tensor is given in therms of the electromagnetic potential $A_\m$ by $F_{\m\n} = \p_\m A_\n-\p_\n A_\m$. \\
A well known solution of these equations (\ref{field-eq1})-\eqref{field-eq2} is given by the rotating C-metric \cite{teo}, a subclass of the Plebanski-Demianski family  \cite{griffiths}. It describes a (dyonically) charged and rotating black hole which is accelerating along the axis of symmetry under the action of a string-like (or strut-like)  force \cite{podolsky}. In the limit of vanishing acceleration, $A \rightarrow 0$,  this spacetime exactly reduce to the standard Kerr-Newman (KN) black hole. It is convenient to parametrise the accelerating metric in the following form
\bea \label{metric}
            ds^2 &=& \frac{1}{(1+\tilde{r} x A)^2} \left\{ \frac{G(\tilde{r})}{\tilde{r}^2+a^2x^2} \left[ d\tilde{t} + a (1-x^2) \D_\varphi d \tilde{\varphi}  \right]^2 - \frac{\tilde{r}^2+a^2 x^2}{G(\tilde{r})} d\tilde{r}^2    \right. \nn \\
        & & \hspace{1.8cm}   + \left.  \frac{H(x)}{\tilde{r}^2+a^2x^2} \left[ (\tilde{r}^2+a^2) \D_\varphi d \tilde{\varphi} + a d\tilde{t}  \right]^2  + \frac{\tilde{r}^2+a^2 x^2}{H(x)} dx^2 \right\} \quad ,
\eea
where\footnote{This solution holds also in presence of the cosmological constant, just upgrading $G(\tilde{r})$ with $G_\L(\tilde{r})=G(\tilde{r})+\frac{\L}{3}\left( \tilde{r}^4+\frac{a^2}{A^2}\right)$. In this case the horizon structure becomes algebraically more involved, moreover they do not coincide any more with the Kerr-Newman-(A)dS ones, because of the explicit dependence on the accelerating parameter $A$. Of course the action (\ref{EM-action}) end equations of motion (\ref{field-eq1}) also have be properly modified to include the cosmological constant.}
\bea \label{G(r)}
         G(\tilde{r})  &:=& \left( A^2 \tilde{r}^2 -1\right) \left( \tilde{r} - r_+\right) \left( \tilde{r} - r_- \right)  \quad ,\label{Gr}\\
     \label{H(x)}    H(x) &:=&  \left( 1 -x^2\right) \left( 1 + A x r_+\right) \left( 1+ A xr_- \right)  \quad .
\eea 

While the electromagnetic potential remains basically the same of the (non-accelerating) Kerr-Newman solution
\beq \label{Apot}
            A_\m = \left\{ - \frac{q \tilde{r} + p a x }{\tilde{r}^2+a^2 x^2} , \ 0  , \  0 , \ - \frac{a q \tilde{r} (1-x^2) - p x (\tilde{r}^2 + a^2) }{\tilde{r}^2+a^2x^2} \D_\varphi \right\} \quad ,
\eeq

The real constants $ m  , \  a , \  A , \ q \ $ and $p$ respectively parametrise  the mass, angular momentum (for unit mass), the acceleration, the electric and magnetic charge of the black hole, but they coincide with these latter quantities only in the limit of vanishing acceleration $A \rightarrow 0$.\\
From the weak field limit, that is $m=a=q=p=0$, the parameter $A$ can be clearly interpreted as the uniform acceleration felt by a test particle at the origin $\tilde{r}=0$ \cite{podolsky}, \cite{acc-string}.
Generally accelerating black holes have two asymmetrical nodal singularities on the poles (located at $x=\pm1$), proportional to
\beq
           \lim_{x\rightarrow \pm 1} \frac{2\pi}{(1-x^2)} \sqrt{\frac{g_{\tilde{\varphi}\tilde{\varphi}}}{g_{xx}}}   = 2\pi \D_\varphi \left(1 \pm A r_+ \right) \left(1 \pm A r_- \right)                              \quad.
\eeq
One of these conical singularity can be easily removed by rescaling the range of the azimuthal coordinate $\varphi$, or equivalently, as in our case, by introducing a constant coefficient $\D_\varphi$ to keep the $\varphi$ range $[0,2\pi]$. For instance, imposing the regularity on the north pole ($x=1$), we set
\beq
           2\pi \D_\varphi \left(1+ A r_+ \right) \left(1 + A r_- \right) = 2 \pi  \qquad \Rightarrow \qquad \D_\varphi = \frac{1}{1+ 2 m A + A^2 (a^2 + q^2  + p^2)} \quad .
\eeq
But, because of the asymmetric conicity, in order to remove also the second angular deficit (or excess) and to remain with a full regular metric, an extra parameter is needed, such as the intensity of an external electromagnetic field. We will study this regularisation in section \ref{magn-acc}. The coordinate $x$ is related to the usual polar angle by $x = \cos \theta$, so its range is $x\in [-1,1] $.\\ 
The position of the horizons can be obtained as the zeros of the $G(\tilde{r})$ function (\ref{Gr}). As on the KN metric, the inner and outer horizons, $\tilde{r}=r_\pm$, are located at 
\beq \label{rpm}
                 r_\pm = m \pm \sqrt{m^2-a^2-q^2-p^2} \quad .
\eeq
For $r=1/A$ we encounter an accelerating horizon, $r_A$, which is supposed to lay beyond the event horizon $r_+$, hence constraining the range of parameters such that $A^{-1}> m + \sqrt{m^2-a^2-q^2-p^2}$.  The black hole become extremal when the inner and outer horizon coincides, for $m=\sqrt{a^2+q^2+p^2}$, at radial distance $r_e = \sqrt{a^2+q^2+p^2}$. On the other hand the extremality condition is not directly affecting the position of the accelerating horizon $r_A$.\\
The black hole area is given by 
\beq \label{area-coni}
                \mathcal{A} = \int_0^{2\pi} d\tilde{\varphi} \int_{-1}^1 dx  \sqrt{g_{\tilde{\varphi}\tilde{\varphi}}g_{xx}} \ \bigg|_{\tilde{r}=r_+} = 4 \pi \D_\varphi \frac{r^2_+ +a^2}{1-A^2 r_+^2}   \quad .
\eeq
The null acceleration limit for the solution (\ref{metric})-(\ref{Apot}), corresponding to $A \rightarrow 0$,  is well defined and gives the standard Kerr-Newman spacetime.\\
In the following it will be useful to know the angular velocity $\Omega_J$ and the Coulomb electromagnetic potential  $\Phi_e$ of the horizon respectively given by
\beq
                   \Omega_J :=  -\frac{g_{\tilde{t} \tilde{\varphi}}}{g_{\tilde{\varphi}\tilde{\varphi}}} \ \Bigg|_{\tilde{r}=r_+} = - \frac{a}{a^2+r_+^2} \ \frac{1}{\D_\varphi} \quad ,
\eeq
and
\beq \label{coul-pot}
              \Phi_e := - \chi^\m A_\m\bigg|_{\tilde{r}=r_+}  =  \frac{q \ r_+}{a^2 +r_+^2}  \quad ,
\eeq
where $\chi=\p_{\tilde{t}} + \Omega_J\p_{\tilde{\varphi}} $ was considered as Killing vector generating the event horizon.
Their extremal limits, for $r_+ \rightarrow r_e $, will be called  $\Omega_J^{ext}$ and $\Phi_e^{ext}$, while $\D_\varphi^{ext}$ is defined as $ \lim_{r_+ \rightarrow r_e} \D_\varphi $.\\
Note that $  \Omega_J $ and $\Phi_e$ depend on the choice of the coordinate, for instance the factor $\D_\varphi$ can be absorbed in a new time coordinate to get the same angular velocity on Kerr-Newman on the horizon. So also the surface gravity or the temperature, may appear different. Nevertheless other choices of the frame will affect also the near horizon geometry (in particular in this trough $\k$), compensating these apparent local discrepancy. Frame independent quantities can be defined thanks to integrability and are presented in the next subsection (\ref{barT})-(\ref{barOm}).    \\
The electric and magnetic charges remain basically the same of the Kerr-Newman black hole, up to the factor $\D_\varphi$
\beq \label{Qq}
        \mathcal{Q} =\frac{1}{8\pi} \int_\mathcal{S} F^{\m\n} dS_{\m\n} = - \frac{1}{4\pi} \int_0^{2\pi} d\tilde{\varphi}  \int_{-1}^1 dx \ \sqrt{g_\mathcal{S}} \ n_\m \s_\n F^{\m\n}  =  q \ \D_\varphi   \quad ,
\eeq
\beq
      \mathcal{P} = \frac{1}{4\pi} \int_\mathcal{S} F_{\m\n} \ dx^\m \wedge dx^\n =  p \ \D_\varphi   \quad ,
\eeq
where $ dS_{\a\b} = - 2 n_{[\a}\s_{\b]} \sqrt{g_\mathcal{S}} \ d\tilde{\varphi} dx $ and $ \sqrt{g_\mathcal{S}} = \sqrt{g_{xx} g_{\tilde{\varphi} \tilde{\varphi}}} $ defines the two-dimensional volume element of the integration surface $\mathcal{S}_t$, surrounding the black hole event horizon at fixed time and fixed radial distance. We also defined $n_\m$ and $\s_\n$ as the two orthonormal vectors, respectively time-like and space-like, normal to the surface $ \mathcal{S}_t $.  \\
Similarly, defining the rotational Killing vector  $\xi^\m_{(\varphi)}=\p_{\tilde{\varphi}}$, we obtain the following value for the angular momentum \cite{gao}
\beq \label{J}
       \mathcal{J} = \frac{1}{16\pi}  \int_ {\mathcal{S}_t} \left[ \nabla^\a \xi^\b_{(\varphi)} + 2 F^{\a\b} \xi^\m_{(\varphi)} A_\m \right] dS_{\a\b}  = a m \D_\varphi^2    \quad .
\eeq

\subsection{Mass and First Law of Thermodynamics for Accelerating Black Holes}
\label{mass-acc}

Computing the mass for accelerating black holes, because their unusual asymptotic, it is a non-trivial task and, up to the author knowledge, it has not achieved at the moment, although some attempts were done recently in \cite{ray} and \cite{greg}. However some hints can come from the thermodynamics. In fact, exploiting some of the results found in \cite{mass-mkn} for a different deformations of the Kerr-Newman black hole, it is possible to find the unique integrable mass coherent with the first law of thermodynamics.\\
When treating with metrics with unconventional asymptotic falloff, a fundamental step in the analysis of the mass consists in the identification of the canonical symmetry associated with the energy, which in general is not $\p_t$ as it occurs in case of asymptotic flatness, for the standard Kerr-Newman solution. Just consider the Kerr-AdS spacetime \cite{claudio} for a well known counterexample, where the normalisation if the Killing vector $\p_t$ is fixed by the asymptotic symmetry algebra. Naive election of this normalisation gives masses that naturally does not fulfil the laws of thermodynamics, unless adjusting adding ad-hoc therms in the first law, as it occurs in \cite{gibbons-pope}, for instance in the case of Kerr-Newman black hole embedded in an external field.\\
In this subsection we consider, for simplicity, the electric charge only in the solution (\ref{metric})-(\ref{Apot}), that means setting $p=0$, and we take as the canonical Killing vector associated to the energy $\a \p_t$, normalised with a integrability factor $\a$, which eventually can be used to define a canonical time $t_{can}=t/\a$. \\ 
Thus the integrable mass continuously connected to the Kerr-Newman one (in the null acceleration limit, $ A \rightarrow 0 $) and obeying to the standard first laws of thermodynamics
\beq \label{first}
        \d \mathcal{M} =  \bar{T} \d \mathcal{S} + \bar{\Omega} \d \mathcal{J} + \bar{\Phi} \d \mathcal{Q} \quad ,
\eeq
        is given by
\beq \label{mass}
        \mathcal{M} = m \frac{\sqrt{1+a^2 A^2} \sqrt{1- A^2 (a^2+q^2) +  2 A \sqrt{m^2-q^2-a^2} \big]}}{\sqrt{1 +  A^2 (a^2+q^2) - 2 A m } \left[ 1 +  A^2 (a^2+q^2) + 2 A m  \right]^{3/2} } \quad .
\eeq

The explicit expression for the normalization factor is 
\beq \label{alfa}
       \a =  \frac{\Big[ a^2 + \left( m + \sqrt{m^2-a^2-q^2} \right)^2 \Big] \left[ 1 - \frac{(q^4 + 4 a^2 m^2) \left[ 1 + A^2 \left( a^2 + q^2 - 2m^2- 2m\sqrt{m^2-a^2-q^2} \right) \right]^2 }{\left[ 1 + A^2 (a^2+q^2) + 2 m A \right]^2 \left(q^2 - 2m^2 - 2m \sqrt{m^2 -a^2 - q^2}  \right)^2} \right] }{4 \mathcal{M} \sqrt{m^2-a^2-q^2} \left[ 1- A^2 \left( m + \sqrt{m^2-q^2-a^2} \right)^2 \right] } \quad .
\eeq 

The frame independent thermodynamic potential $\bar{T}, \bar{\Om} , \bar{\Phi}$ are defined as
\bea
     \label{barT}  \bar{T} &:=& \a T_H \ = \ \frac{1}{8\pi \mathcal{M}} \left[ 1 - \frac{\pi^2}{\mathcal{S}^2} \left( 4 \mathcal{J}^2 + \mathcal{Q}^4 \right)    \right] \  \equiv \  \frac{\p \mathcal{M}}{ \p \mathcal{S}} \bigg|_{\mathcal{JQ}}
          \hspace{0.2cm}    ,   \\
     \label{barOm}   \bar{\Om} &:=& \a (\Om_J-\Om_{int}) \ = \ \frac{\pi \mathcal{J}}{\mathcal{M} \mathcal{S}} \  \equiv \  \frac{\p \mathcal{M}}{ \p \mathcal{J}} \bigg|_{\mathcal{SQ}} \hspace{2.5cm}  ,  \\
     \label{barPhi} \bar{\Phi} &:=& \a (\Phi_e-\Phi_{int}) \ = \ \frac{\pi \mathcal{Q}}{2 \mathcal{M} \mathcal{S}} \left( Q^2 + \frac{\mathcal{S}}{\pi	} \right) \  \equiv \  \frac{\p \mathcal{M}}{ \p \mathcal{Q}} \bigg|_{\mathcal{SJ}} \hspace{0.55cm}  .
\eea 

where $\Om_{int}$ and $\Phi_{int}$ are also fixed by integrability conditions. But possibly it is easy to choose a gauge for the solution (\ref{metric})-(\ref{Apot}), by properly shifting the electromagnetic potential and the off-diagonal term of the metric by a constant, for which $\Om_{int}$ and $\Phi_{int}$ are null, as explained in \cite{mass-mkn}. These settings, together with the time coordinate normalised by a factor $\a$ and the $\varphi$ angle co-rotating with $\Phi_{int}$, constitute the, so called, canonical frame.\\ 
The Hawking temperature $T_H$ is defined as usual in terms of the surface gravity, the explicit value for the accelerating case can be found in eq (\ref{th}). $S$ refers to the entropy of the black hole and, as usual, is a quarter of its area (\ref{area-coni}) $S=\mathcal{A}/4$. \\
It is easy to verify that the mass (\ref{mass}) also satisfy the Smarr formula
\beq
        \mathcal{M} = 2 \bar{T} \mathcal{S} + 2 \bar{\Om} \mathcal{J} + \bar{\Phi} \mathcal{Q} \quad .
\eeq

We present the details for the non-rotating metric, thus also $a=0$. In this case  the angular momentum $\mathcal{J}$ is null and the mass can be read from (\ref{mass})
\beq
         \mathcal{M}\big|_{{a=0}} = m \frac{\sqrt{1- A^2 q^2 + 2 A \sqrt{m^2-q^2} }}{\sqrt{1+ A^2 q^2 - 2 A m} \left[1+ A^2 q^2 + 2 A m \right]^{3/2} }   \quad .
\eeq
It can be easily checked that (\ref{first}) is satisfied using the coulomb potential (\ref{coul-pot}) and 
$$\Phi_{int}\Big|_{a=0} = \frac{m q A}{A q^2 + \sqrt{m^2-q^2}} \quad .$$  
A different value for the mass is given in \cite{greg}. It is computed using the usual Killing vector $\p_t$, with the normalisation typical of trivial (null curvature) asymptotic, but accelerating black holes are endowed with different asymptotic. In fact using the mass of \cite{greg} the first law of black hole thermodynamics can not be fulfilled in general, but only adding extra constraints on the physical parameters. \\
The mass for the uncharged sub-case is also well defined and it follows smoothly from (\ref{mass}) in the limit $q \rightarrow 0$.\\ More details and a direct computation of the mass is outside the scope of the paper and will be presented elsewhere \cite{marcoa-acc-M}. Anyway note that the spirit of this subsection is to find a mass for the c-metric fulfilling the first law of black hole thermodynamics, which is assumed. However we are aware that the first law might not hold for this kind of singular solutions, because of the conical defects. Therefore in \cite{marcoa-acc-M} we analyse regular accelerating black holes.

\section{Near horizon geometry at extremality}
\label{acc-NH}

In order to analyse the region near the extremal accelerating Kerr-Newman black hole (EAKN) event horizon $r_e$, we follow the usual prescription of \cite{compere-review}, originally developed in \cite{horowitz}. We have to introduce new dimensionless coordinates $(t,r,\varphi)$ defined as follows
\beq \label{change-coordinate}
            \tilde{r}(r) := r_e+ \l r_0 r    \qquad  ,  \qquad      \tilde{t}(t) := \frac{r_0}{\l} t     \qquad  ,    \qquad     \tilde{\varphi}(\varphi,t):= \varphi + \Omega_J^{ext} \ \frac{r_0}{\l} t  \qquad  , \qquad
\eeq
where the constant $r_0$ is brought in to cancel the overall scale of the near-horizon geometry. When an electromagnetic potential $A_\m$  is present, also it is needed a gauge transformation of this kind
\beq \label{elect-shift}
          A_{\tilde{t}} \rightarrow A_{\tilde{t}} + \Phi_e  \quad .
\eeq 
Thus the near horizon,  extreme, accelerating Kerr-Newman geometry (NHEAKN) is obtained as the limit of the  EAKN for $\l \rightarrow 0$. Exactly as in the non-accelerating case, the NHEAKN geometry can be cast in the general form of the near-horizon geometry of spinning extremal black holes, endowed with the $SL(2,\mathbb{R})$ symmetry \cite{Kunduri:2007vf}, which can be expressed as a warped and twisted product of $AdS_2 \times S^2$
\beq \label{metric-near}
           ds^2 = \G(x) \left[ -r^2 dt^2 + \frac{dr^2}{r^2} + \a^2(x) \frac{dx^2}{1-x^2} + \g^2(x) \ \big( d\varphi + \k r dt \big)^2 \right] \quad ,
\eeq 
where  
\bea \label{fields1}
           \G(x) &=&\frac{a^2 x^2  + r_+ r_-}{\left[1-A^2 r_+ r_- \right] \left( 1+A x \sqrt{r_+ r_-} \right)^2}  \qquad   \qquad , \hspace{1.1cm} r_{0} \ = \ \pm \sqrt{\frac{a^2 + r_+ r_-}{1-A^2 r_+ r_-}}   \qquad    \quad ,  \qquad   \\
    \label{fields2}    \g(x)  &=& \pm \frac{(a^2 + r_+ r_-) \sqrt{1-x^2} \D_\varphi^{ext}}{\G \sqrt{1-A^2 r_+ r_-} \left( 1+A x \sqrt{r_+ r_-} \right)}       \quad  \   \ \quad , \qquad \hspace{0.5cm} \k \ = \  - \frac{2 a r_{0}^2\sqrt{r_+ r_-}}{(a^2 + r_+ r_-)^2 \D_\varphi^{ext}} \quad , \\
  \a(x) & = & \pm \frac{\sqrt{1-A^2 r_+ r_-}}{1+xA\sqrt{r_+ r_-}}    \label{fields3}   \qquad . \quad 
\eea
Also the electromagnetic connection fall into the same general class of near horizon gauge potential
\beq \label{near-A}
               A = \ell(x) (d\varphi+\k r dt) - \frac{e}{\k} d\varphi \qquad ,
\eeq
where
\beq\label{near-ell}
           \ell(x) = - \frac{r_0^2}{\k} \frac{q (r_+ r_- - a^2 x^2) + 2 a x p  \sqrt{r_+ r_-}}{(r_+ r_- + a^2 x^2) (a^2 + r_+ r_-) }  \qquad \quad  , \quad  \qquad e = q r_0^2 \frac{r_+ r_- -a^2}{\left(r_+ r_- +a^2 \right)^2} \quad .
\eeq
It is interesting to note that this near-horizon geometry differs from the usual Kerr-Newman ones\footnote{We will refer to the Kerr-Newman quantities with an extra zero pedix.}, which can be easily obtained in the $A \rightarrow0$ limit from (\ref{fields1})-(\ref{near-ell}):
\bea 
\G_0(x) &=& a^2 x^2  + r_+ r_-  \hspace{2.4cm} ,  \hspace{1.05cm}  \a_0(x) \ = \ \pm 1  \hspace{1.5cm}   ,    \label{fields01}   \\
\g_0(x)  &=&  \pm  \frac{(a^2 + r_+ r_- ) \sqrt{1-x^2}}{\G_0(x)} \hspace{0.9cm} ,      \hspace{1.1cm}     r_{0_0}^2 \ = \ a^2 + r_+ r_-    \qquad   ,        \label{fields02}   \\
\k_0 & = &  - \frac{2 a \sqrt{r_+ r_-}}{a^2 + r_+ r_- }    \hspace{2.45cm}  .
\label{fields03}       
\eea
That is not a trivial statement because, as shown in \cite{magn-kerr-cft}, the near-horizon geometry of Kerr-Newman black holes distorted by an external magnetic field, remains, at extremality, isomorphic to the unmagnetised metric, near the horizon. In fact results stating that this is a general behaviour in four-dimensions in standard General Relativity, not only pertinent to external magnetic field deformations, have recently appeared  \cite{li}. However, as we explain below, there is no tension with our
work. But in presence of this kind of acceleration it is easy to show that the near-horizon geometry does not belong to the non-accelerating Kerr-Newman class any more. Indeed  we can perform a coordinate transformation
\beq \label{change-coord}
            x(y)= - \frac{A\sqrt{r_+ r_-} \pm y}{1\pm Ay \sqrt{r_+ r_-}} \quad ,
\eeq
to reabsorb the function $\a$, as in the standard KN case. But then it is clear that the transformed form of $\G[x(y)]$, which reads 
\beq
             \G(y) = \frac{r_+ r_- \left(1-A y \sqrt{r_+ r_-} \right)^2 + a^2 \left(y-A\sqrt{r_+ r_-} \right)^2}{\left[1-A^2(r_+ r_-)\right]^3} \quad ,
\eeq
differs from $\G_0(x)$, because of the linear term in the coordinate $y$,  and we have no extra freedom to make them match.  Note that this feature mainly depends on the source of the acceleration only, the string (or strut); therefore it also holds in absence of the electromagnetic potential ($q=0, p=0$). It can be ascribed to the fact that the cosmic string is producing a conical singularity, hence the spacetime is not regular and a delta source should be added in the energy momentum tensor. Indeed, as will be shown in section \ref{magn-acc}, when the acceleration is generated by an external magnetic field, the metric can be arranged to be completely regular introducing a constraint between the parameters. In that case the matter energy momentum tensor remains only the Maxwell one and as suggested in \cite{kunduri08}, the extremal near horizon geometry corresponds to the one of extremal Kerr-Newman solution. Actually the regularity constrain precisely coincides with the cancellation of the linear term of in $\G$.\\
When the metric is not regularised, the presence of conical singualrity evades the infinitesimal transverse deformations considered in \cite{li}, that is why we have no extremal Kerr near horizon geometry, without any contradiction with the result of \cite{li}. \\
Because the near horizon geometry of the AEKN black hole can be cast in the general form (\ref{metric-near}), its isometry is generated by the usual\footnote{Note that the deformation due to the acceleration enters only in $\k$.} following Killing vectors
\bea 
              \z_{-1} = \p_t  \qquad &,& \qquad \z_0 = t \p_t - r \p_r \qquad \\
              \z_1 = \left( \frac{1}{2r^2} +\frac{t^2}{2} \right) \p_t - t \  r \ \p_r - \frac{\k}{r} \p_\varphi \qquad  &,& \qquad L_0 = \p_\varphi \qquad .          
\eea
From their non null commutation relations 
\bea
              [\z_0 , \z_{\pm}] = \pm \z_{\pm}   \qquad \quad , \quad \qquad [ \z_{-1} , \z_1 ] = \z_0 \qquad       
\eea  
we understand that they span the  $SL(2,\mathbb{R}) \times U(1)$ algebra, where $L_0$ generate the $U(1)$ algebra. The generators of the  infinitesimal isometries are normalised to simplify the commutation rules.\\
Therefore the presence of the acceleration is not spoiling the near horizon symmetry of the non accelerating case, at least in the extremal case, which is a key point in the formulation of the Kerr/CFT correspondence.\\
Note that while the near horizon geometry (\ref{metric-near}) is a characteristic of the event horizon, other killing horizons such as, for examples, the accelerating horizon $r_A$, can not be expressed as a warped product of $AdS_2 \times S^2$, neither in the extremal case.\\
According to the Kerr/CFT correspondence it is possible to infer the thermodynamic properties of extremal black holes from the asymptotic symmetry of their near horizon fields. 
Thus specification of proper boundary conditions, for the near horizon metric (\ref{metric-near})-(\ref{fields3}) and electromagnetic potential (\ref{near-A})-(\ref{near-ell}), become necessary. We will borrow the usual boundary conditions for the theory we are considering \cite{stro-duals} - \cite{compere-review}: the fall-off behaviour for the metric, at large radial distance $r$, is taken as follows  
\begin{align}
g_{tt}& = \mathcal{O}\left({r^2}\right) \ \ , \hspace{0.7 cm} g_{t\varphi}= \k \ \G (x) \ \g^2(x) \ r + \mathcal{O}\left({1}\right),\nn\\
g_{tx} & = \mathcal{O}\left({1\over r}\right)\ ,\qquad g_{t r} = \mathcal{O} \left({1\over r^2}\right) , \hspace{0.8cm} g_{\varphi \varphi} = \mathcal{O}(1) \ , \nn \\
g_{\varphi x} & = \mathcal{O}\left({1\over r}\right) \  , \qquad g_{\varphi r}= \mathcal{O} \left({1\over r}\right) \ \ , \qquad g_{x r} = \mathcal{O} \left({1\over r^2}\right) , \label{bound-cond} \\
g_{xx}&= \frac{\Gamma(x)\alpha(x)^2}{1-x^2} + \mathcal{O} \left({1\over r}\right) \ , \hspace{1.9cm} 
g_{rr}= \frac{\Gamma(x)}{r^2}+\mathcal{O} \left({1\over r^3}\right),\nn
\end{align}
while the electromagnetic field is considered to decay in the following way
\begin{align}
	A_{t}&= \mathcal{O}\left({r}\right) \qquad , \qquad A_{\varphi}= \ell(\theta) - \frac{e}{\k} + \mathcal{O}\left({\frac{1}{r}}\right),\nn\\
	A_{x} &= \mathcal{O}\left(1\right) \qquad, \qquad A_{r}= \mathcal{O} \left({1\over r^2}\right) \quad .
\end{align}
These boundary conditions\footnote{In \cite{stro-duals} these boundary condition were supplemented by the zero energy and electric charge excitation condition, i.e. $\d\mathcal{Q}_{\p_t}=0$ and $\d\mathcal{Q}=0$, respectively.} are preserved by the following asymptotic Killing vectors 
\bea 
        \z_\epsilon &=&  \epsilon(\varphi) \p_\varphi -r \epsilon'(\varphi)  \p_r \ \   + \ \ \  \textrm{subleading terms} \quad  , \quad  \label{zep}\\
        \xi_\epsilon &=& -\left[\ell(\theta)-\frac{e}{\k} \right]  \epsilon(\varphi ) \ \  + \ \ \ \textrm{subleading terms}  \quad . \quad    \label{xiep}
\eea
On the bulk the boundary conditions (\ref{bound-cond})-(\ref{xiep}) are preserved also by some of the near horizon symmetry generators: ${\z_{-1}, \z_0}$, but not by $\z_1$. Expanding the generators (\ref{zep}) - (\ref{xiep}) in Fourier modes such that 
\beq
            \e(\phi) = -e^{-i n \phi} \quad ,
\eeq            
we can verify that each $m-$mode couple in the Fourier series expansion can be considered as a generator, $L_m = (\z_m,\xi_m)$, which obey the following  Witt algebra (Virasoro algebra without the central extension) 
\beq
                 i \  [L_m,L_n] = (m-n) \ L_{m+n} \quad .
\eeq
The commutation bracket are defined by
\beq \label{dewitt}
             [L_m,L_n] := [( \z_m , \xi_m ) , ( \z_n , \xi_n )] = ([\z_m,\z_n],[\xi_m,\xi_n]_\z) \quad , 
\eeq
where $[\z_m,\z_n]$ is the standard Lie commutator, while $[\xi_m,\xi_n]_\z := \z^\m_m \p_\m \xi_n - \z^\m_n \p_\m \xi_m $ . \\

\section{Microscopic Entropy}
\label{micro}

The emergence of the Witt asymptotic algebra inspires the hypothesis that some quantum gravity features of the near horizon region of the accelerating extremal Kerr-Newman black hole can be deduced from a dual two-dimensional CFT living on the KHAEKN boundary. \\
Evaluating the Dirac bracket between the charges associated with the generators of the asymptotic symmetries (\ref{zep}) - (\ref{xiep}), one can observe that the Witt algebra is enlarged into the full Virasoro algebra, with a non-null central extension. The central charge can be calculated as the coefficient of the cubic factor in the $m$-expansion of the following asymptotic charge  

\beq
                c_J = 12 \ i \lim_{r \rightarrow \infty} \mathcal{Q}_{L_m} ^{\mathrm{Einstein}} [\mathcal{L}_{L_{-m}} \bar{g}; \bar{g}] \Big|_{m^3} \quad ,
\eeq

where $\mathcal{L}_{L_{-m}} \bar{g}$ is the Lie derivative of the background metric along the generator $L_{-m}$ and the fundamental charge formula for is given, for general relativity \cite{35}, by

\beq
\mathcal{Q}_{L_{m}}^{\mathrm{Einstein}} [h ; \bar{g}] = \frac{1}{8 \pi G_N}\int_S dS_{\mu\nu}\Big( \xi^\nu \nabla^\mu h +\xi^\mu \nabla_\sigma h^{\sigma\nu}+\xi_\sigma \nabla^\nu h^{\sigma\mu} +\frac{1}{2}h \nabla^\nu \xi^\mu +\frac{1}{2}h^{\mu\sigma}\nabla_\sigma \xi^\nu+\frac{1}{2}h^{\nu\sigma}\nabla^\mu \xi_\sigma\Big)\ . \nn
\eeq

Here $h$ is defined as $h:=\bar{g}^{\m\n} h_{\m\n}$, $\mathcal{Q}_{L_{m}}^{\mathrm{Einstein}} [h ; \bar{g}]$ represents the conserved charge associated with the Killing vector $\xi^\m$  of the linearised metric $h_{\m\n}$ around the background $\bar{g}_{\m\n}$, while $\mathcal{S}$ and $dS_{\m\n}$ are defined in section \ref{acc-rev}. From the near horizon geometry (\ref{metric-near}) we obtain a general expression for the central charge given by

\beq \label{cc}
         c_J = 3 \k \int_{-1}^1 \frac{dx}{\sqrt{1-x^2}} \ \G(x) \a(x) \g(x)  \quad .
\eeq

Note that matter does not affect directly the value of the central charge but it enters only implicitly through the constant $\k$ and the functions $ \G(x) , \g(x) , \a(x)  $. This is not a surprise but a typical behaviour for the theory (\ref{EM-action}) and class of near horizon geometry (\ref{metric-near}) we are considering here, as shown in \cite{compere-review}. \\
Note also that the central charge does not depend on the particular choice of the boundary conditions, but only on their existence.\\
Then making use of the explicit form of the fields of the NHEAKN metric we can evaluate the central charge for the near horizon geometry of the accelerating extremal Kerr-Newman black hole

\beq \label{cca}
         c_J = \frac{12 a \sqrt{r_+ r_-}}{\left[ 1 - A^2 r_+ r_-\right]^2}\quad .
\eeq 

Therefore the Witt algebra (\ref{dewitt}) acquires a central extension, becoming a Virasoro algebra

\beq
              [ \mathcal{L}_m,\mathcal{L}_n] = (m-n) \mathcal{L}_{m+n} + \frac{c_J}{12} m (m^2-B) \d_{m,-n}
\eeq

Where the real parameter $B$ is a trivial central extension that can be put to 1 by shifting the background value of the charge $\mathcal{L}_0$.\\
The framework of the Kerr/CFT correspondence exploits the assumption that near horizon geometry of extremal black holes can be described by the left sector of a CFT in two dimensions. For these latter theories Cardy found \cite{cardy}that the asymptotic grown of states density, in the microcanonical ensamble, is given by\footnote{The right degrees of freedom are neglected, because we are considering only the extremal case.}

\beq \label{cardy0}
          \mathcal{S}_{CFT} = 2 \pi \sqrt{\frac{c_L \mathcal{L}_0}{6}} \quad ,
\eeq      

thus it depends only on the central charge of the theory and the zero eigenvalue $\mathcal{L}_0$. This formula (\ref{cardy0}) is valid for unitary and modular invariant CFTs and for $\mathcal{L}_0 >> c_L$. Using the definition of left temperature

\beq
       \frac{\p \mathcal{S}_{CFT}}{\p \mathcal{L}_0 } = \frac{1}{T_L}
\eeq

it is possible to transform the (left sector of the) Cardy formula in the canonical ensemble to get

\beq \label{cardy-can-ens}
            \mathcal{S}_{CFT} = \frac{\pi}{3} c_L \ T_L \quad .
\eeq

In this setting the validity of (\ref{cardy-can-ens}) can be quantified by asking large temperatures $T_L>>1$, which imply large number of excited degrees of freedom.\\
Since, in the extremal case, we are dealing with the rotational excitations around $\p_\phi$, we have the presence of the left sector quantities only. We cannot associate to the left temperature the Hawking temperature $T_H$ because, even though it is directly affected by acceleration, at extremality it vanishes on the event horizon, as the surface gravity $k_s$, because the outer and inner horizon overlap in a double degenerate horizon

\beq \label{th} 
            T_H  :=  \frac{ \hbar \ k_s}{ k_B \ 2 \pi} =  \frac{1}{2 \pi} \sqrt{-\mezzo \nabla_\m \chi_\n \nabla^\m \chi^\n} \ \ \Bigg|_{r_+} = \frac{1-A^2 r_+^2}{2 \pi} \ \frac{r_+ - r_-}{2 (r_+^2+a^2)} \qquad .
\eeq

Therefore, to take into account the rotational degrees of freedom, the Frolov-Thorne vacuum is used to define a temperature. This can be considered as a generalisation of the Hartle-Hawking vacuum originally built for defining the Hawking temperature for the static Schwarzschild black hole. The Frolov-Thorne vacuum is defined for stationary black holes, in the region where a timelike Killing vector, such as the generator of the horizon, remains timelike. At least it occurs in the proximity of the horizon. The Frolov-Thorne temperature is a geometric quantity, which depends on the metric and matter field, but not straightly on the theory. At extremality it is defined as 

\beq \label{Tphi}
            T_\varphi :=   \lim_{\tilde{r}_+\rightarrow \tilde{r_e}}  \frac{T_H}{\Omega_J^{ext}-\Omega_J}  = - \frac{\D_\varphi^{ext}}{4 \pi} \frac{(a^2 + r_+ r_-) \left[ 1-A^2 r_+ r_- \right]}{a \sqrt{r_+ r_-}}    = \frac{1}{2 \pi \k } \qquad .
\eeq

It can be considered as the vacuum state for spinning or charged extreme black holes.\\
Finally inserting the central charge (\ref{cc}) and the rotational left temperature (\ref{Tphi}) in the Cardy formula (\ref{cardy-can-ens}) we can obtain the value of the entropy of the conformal field theory model associated to the extremal accelerating black hole 

\beq \label{entropy}
           \mathcal{S_{CFT}}=\frac{\pi^2}{3} c_L T_L = \frac{ \pi (a^2 + r_+ r_-) \D_\varphi^{ext} }{1-A^2 r_+ r_-} = \frac{1}{4} \mathcal{A}^{ext} \quad .
\eeq

Note that this dual entropy precisely coincides with the classical Bekenstein-Hawking entropy of the black hole, i.e. with one quarter of its event horizon area, as expected.\\
It is interesting to point out also that the presence of the extra parameter due to the acceleration $A$ it is not improving the applicability of the Cardy formula, with respect to the standard $A=0$ case, because it does not affect the possibility of having a large temperature $ T_L >> 1 $. On the contrary it was shown in\cite{micro-RN} and \cite{magn-kerr-cft} that the presence of an extra parameter related to the external magnetic field improves the plausibility of the Cardy formula application, since it allows to enlarge the temperature,  for some range of the parameters. That's a further motivation to consider, in the next section, these external electromagnetic fields as regulators.  \\
On the other hand the limits to the non accelerating standard case $A\rightarrow0$ are well defined on any step, so the standard Kerr/CFT is easily and clearly recovered as a subcase.\\
The presence of an Abelian gauge field, given by the Maxwell electromagnetic connection $A_\m$, makes available also an alternative CFT dual picture. In fact, instead of using the rotational symmetry around the azimuthal axis, we can take advantages of the $U(1)$ symmetry of the electromagnetic potential through a Kaluza-Klein uplift in five dimensions. Thus the Abelian gauge field is thought to be wrapped around a compact extra dimension $\psi$, with period $2\pi R_\psi$, which define a killing orbit $\p_\psi$. A chemical potential associated with the direction generated by $\p_\psi$ can be defined as explained in \cite{compere-review}. In that case the Frolov-Thorne temperature is given in units of $R_\psi$ by
\beq
       T_\psi = T_e R_\psi \quad .
\eeq
In analogy with the rotational picture, the electric chemical potential is defined, at extremality, as

\beq
       T_e := \lim_{r_+ \rightarrow r_e}  \frac{T_H}{\Phi_e^{ext} -\Phi_e} = \frac{(2a^2+p^2+q^2)\left[ 1 - A^2 (a^2+p^2+q^2) \right] }{2 \pi q (p^2+q^2)} = \frac{1}{2 \pi e}\quad .
\eeq
The fact that $T_e$ can be expressed, as in the last equality, in terms of the near-horizon quantity $e$ (\ref{near-ell}), also in this accelerating case, it is a not trivial feature. 
 Hence the temperature associated with the second CFT picture becomes
\beq
          T_\psi = \frac{R_\psi}{2\pi e} \quad .
\eeq 
Assuming, as in the standard Kerr/CFT formulation, that in the extremal case there are no right excitations modes in the conformal model, $T_\psi$ can be considered as the left temperature
\beq\label{Tpsi}
        T_L = T_\psi  \qquad \qquad , \qquad  \qquad    T_R = 0 \quad .
\eeq
Thanks to the five-dimensional uplift the central charge can be computed in a similar way with respect to $c_J$. It is given by
\beq \label{ccQ}
         c_Q = \frac{3 e }{R_\psi} \int_{-1}^1 \frac{\G(x) \a(x) \g(x)}{\sqrt{1-x^2}} \  dx \  = \frac{6 q (q^2 +p^2) \D_\varphi }{\left[ 1 - A^2 (a^2+q^2+p^2) \right]^2 R_\psi} \quad .
\eeq
Finally the entropy of the alternative conformal model dual to the accelerating Kerr-Newman black hole can be written thanks to the Cardy formula (\ref{cardy-can-ens}) and (\ref{Tpsi})-(\ref{ccQ})

\beq \label{entro-alt}
         \mathcal{S_{CFT}} = \frac{\pi^2}{3} \ c_Q T_\psi   = \frac{1}{4} \mathcal{A}^{ext} \quad .
\eeq
Again the entropy of this second dual conformal system coincides with the usual Bekenstein-Hawking gravitational entropy, as in (\ref{entropy}), which corresponds to a quarter of the event horizon area.\\
The main advantage of this second dual picture basically rely in the fact the Kerr/CFT correspondence can be applied even in the lack of rotation (that is for the charged C-metric, when $a=0$). \\
Generalisation in the presence of cosmological constant can be also done directly.\\

The Kerr/CFT formalism might hold also outside the extremal limit, but at the price of adding some ad-hoc extra assumptions on the nature of the central charges. For instance in the standard case of Kerr-Newman one has to assume that the left and right central charges coincides. Moreover it is assumed that the central charges do not change their form (but they change their value) with respect to the extremal case, basically it means that $ c_L = c_R = 12 J $. In practice these values are chosen to match the black hole entropy, so it is not considered satisfactory by some authors \cite{compere-review}. \\ 
On the other hand, even though away from extremality the near horizon geometry looses the $AdS_2$ symmetry, it is still possible to extract some hidden conformal invariance. In fact the equation governing the dynamics of a probe scalar field in the vicinity of the black hole horizon manifests the $ SL(2,\mathbb{R}) \times SL(2,\mathbb{R})$ invariance in a specific low energy regime. This conformal symmetry usually makes possible to compute the left and right-moving temperatures of a CFT model dual to the non-extremal black hole.\\
In presence of the acceleration it is known \cite{separability} that the Klein-Gordon equation  for a probe scalar field, of charge $q_e$, in the rotating C-metric (\ref{metric}) non-backreacting background
\beq  \label{KG}
           ( D^\n D_\n + \m^2 ) \ \Psi(\tilde{t} , \tilde{r} , x , \tilde{\varphi}) = 0 \quad ,
\eeq
is separable only in the massless case, $\m=0$. The covariant derivative $D_\m$ is defined by $D_\n \Psi = \nabla_\n \Psi - i q_e A_\n $. In order to show the decoupling of (\ref{KG})  in a radial and an angular part is convenient to expand the scalar field as 

\beq
      \Psi(\tilde{t} , \tilde{r} , x , \tilde{\varphi}) = (1 + A \tilde{r} x) \  e^{- w_0 \tilde{t} + m_0 \tilde{\varphi} } \ X(x) \ Y(\tilde{r})  \quad ,
\eeq 

where $w_0$ and $m_0$ are the wave frequency and the azimuthal separation constant respectively. The radial scalar field equation becomes (for null magnetic charge, $p=0$)

\beq
\left\{     \partial_{\tilde{r}} \big[ G(\tilde{r}) \p_{\tilde{r}} \big] + \frac{\left[ \frac{a m_0}{\D_\phi} - q_e q \tilde{r} + w_0 (a^2 +\tilde{r}^2) \right]^2}{G(\tilde{r})} + A^2 \tilde{r}(\tilde{r}-m) - C_\ell \right\} Y(\tilde{r}) = 0 \quad ,
\eeq

where $C_\ell$ is the separation constant.\\
The decoupled scalar field equations can be simplified when considered for a specific range of the parameters, that is when the scalar wave has low energy, low mass and low electric charge with respect to the black hole charges. This limit identify the so called ``near region'' of the spacetime, which has not to be confused with the near horizon region of the previous section.\\
In this regime, passing to ``conformal" coordinate, it is possible to exploit the $SL(2,\mathbb{R}) \times SL(2,\mathbb{R}) $ symmetries of the scalar wave equation to obtain a left and right temperature for the conformal model.\\
In presence of the acceleration the assumption about the central charges to remain $c_L=c_R=12J$ is not in general true. Insisting with this assumption constraints the period of the azimuthal coordinate. In fact considering the value of the angular momentum (\ref{J}) we have that $c_L = 12 a m \D_\varphi $, which at extremality coincides with the central charge (\ref{cca}) only if 

\beq \label{costr-D}
        \D_\varphi = \frac{1}{1-A^2 r_+ r_- }   \quad .
\eeq

Note that this is not the value which remove one of the axial nodal singularities. Moreover is not clear how to implement this constraint away from extremality. Therefore away from the extremal case the presence of acceleration raises new issues on an already unsatisfactory picture. A solid approach would consist in an independent computation of the central charge in the non extremal case. This would be very interesting even in the standard case of null acceleration, i.e. the Kerr-Newman case, but, up to the author knowledge, at the moment is not clear how to pursue it.


\section{Regular case: Accelerating and rotating black hole \\in an external magnetic field}
\label{magn-acc}

In this section we want to show, with simple but non-trivial example, that the treatment developed in the previous sections, about the CFT duals of accelerating black holes, can be generalised also when the  the conical singularity, typical of these accelerating spacetimes, is regularised. This can be achieved by means of an external field, still remaining in the realm of the Einstein-Maxwell theory described by the action (\ref{EM-action}). As discovered by Ernst in \cite{ernst-remove}, it is possible to remove the nodal singularity of the C-metric introducing an electromagnetic field of the kind of the Melvin Universe \cite{melvin}. In practice it can be realised by applying an Harrison transformation to the singular electrovacuum solution, at the price of modifying the asymptotic behaviour. From a physical point of view it means that the acceleration is provided by the external electromagnetic field, in spite of the singular string (or strut). This kind of solution were popular some years ago to describe the pair creation of black holes pairs in a external electromagnetic background \cite{gibbons-pair}, \cite{strominger-pair}, \cite{hawking-pair} and recently extended to the rotating case in \cite{rot-pair}. \\
In particular, here, we will focus on a rotating generalisation of the Ernst metric \cite{ernst-remove}, first described in \cite{rot-pair}. In fact this kind of solutions connect the accelerating Demianski-Plebanski family with the magnetised Ernst ones. Basically they describe an accelerating and dyonically charged black hole embedded in an external magnetic universe. Thus the Ernst metric \cite{ernst-remove} can be obtained by tacking the limit of vanishing electric charge, i.e. $ q \rightarrow 0 $, while the Reissner-Nordstrom black in the external magnetic background \cite{ernst-bh-magnetic} be recovered for null acceleration $A \rightarrow 0 $. In practice to obtain this solution an Harrison transformation is applied to the spacetime (\ref{metric})-(\ref{Apot}),  where we set $a=0$ for simplicity. A non-trivial feature of this metric consists in the fact that the accelerating RN spacetime is not static any more, although we have vanished the Kerr rotational parameter $a$. That's because of the Lorentz-like interaction between the intrinsic electric monopole charge of the black hole and the external magnetic field.

The resulting metric and electromagnetic potential, as explained in \cite{rot-pair}, can be written as follows

\beq  \label{elctr-ernst-g}
          ds^2 =         \frac{\left|\L(\tilde{r},x)\right|^2}{(1+A \tilde{r} x)^2} \left[ - \frac{G(\tilde{r})}{\tilde{r}^2} dt^2 + \frac{\tilde{r}^2 d\tilde{r}^2}{G(\tilde{r})}  + \frac{\tilde{r}^2 dx^2}{H(x)}  \right] + \frac{\tilde{r}^2 H(x) (\D_\varphi d\varphi - \om(\tilde{r},x) d\tilde{t} )^2}{(1+A\tilde{r}x)^2 \left|\L(\tilde{r},x)\right|^2 }                             \quad ,
\eeq

\beq \label{elctr-ernst-A}
          A_\m = \big[ A_t( \tilde{r} , x ) , \ 0 , \  0 , \ A_\varphi ( \tilde{r} , x ) \big]    \quad ,
\eeq

where

\bea 
       \L(\tilde{r},x) &=&  1 + B x ( p - i q ) +  \frac{B^2}{4} \left[ \frac{\tilde{r}^2 H(x)}{(1+A \tilde{r} x)^2} + (p^2 + q^2) x^2  \right] \quad ,\\
      \om(\tilde{r},\tilde{t})&=& - \frac{2 q B}{\tilde{r}} +  \frac{q B^3 \left[ (\tilde{r}^2-2m\tilde{r})(1+2A\tilde{r}x + x^2) + x^2 (p^2 + q^2) (1- A^2 \tilde{r}^2) \right] }{2 \tilde{r} \ (1 + A \tilde{r} x)^2} \quad , \label{om} \\
       A_{\tilde{\varphi}} (\tilde{r},x) &=& + \frac{\left[ 2 ( \textrm{Re}(\L) -1 \big) -Bxp\right]  \textrm{Re} (\L) + \left[ \textrm{Im} (\L) \right]^2}{B\left|\L(\tilde{r},x)\right|^2  } \quad ,\\
        A_{\tilde{t}}(\tilde{r},x) &=&  \frac{2 q}{\tilde{r}} + \om(\tilde{r},x) \left[\frac{3}{2 B} -A_{\tilde{\varphi}}(\tilde{r},x) \right] \quad . \label{At}
\eea

$G(\tilde{r}) , \ H(x)$ and $r_\pm$ are defined as in (\ref{G(r)}), (\ref{H(x)}) and (\ref{rpm}) respectively, but now $a=0$. The solution (\ref{elctr-ernst-g})-(\ref{At}) presents nodal singularities on the symmetry axis, as it can be seen by considering a small circle, for fixed time and radial coordinates, around the two semi-axises  $ x = \pm 1 $

\beq \label{con-sin}
      \frac{\textrm{circunference}}{\textrm{radius}} = \lim_{x \rightarrow \pm 1} \frac{2 \pi}{1-x^2} \sqrt{\frac{g_{\tilde{\varphi} \tilde{\varphi}}}{g_{xx} }} =  \frac{32 \pi \D_\varphi \left[1 \pm 2 A m + A^2 (p^2 + q^2) \right]}{ ( \pm 2 + B p )^4 + 2 B^2 q^2 \left[12 + B p (\pm 4 + B p )\right]}
\eeq

Note that these deficit or, depending on parameters, excess angle is asymmetric on the two different hemispheres. Therefore it is possible to remove only one of the conical singularities at a time, let's say we chose to regularise the one on the semi-axis $x=1$, by setting $\D_\varphi$ to 

\beq \label{reg-const}
           \bar{\D}_\varphi = \frac{ ( 2 + B p )^4 + 2 B^2 q^2 \left[12 + B p ( 4 + B p )\right]}{16 \left[1+ 2 A m + A^2 (p^2 + q^2) \right]} \quad .
\eeq 

Now the presence of the external electromagnetic field plays a fundamental role. Because it makes possible, at the same time, the elimination also of the second conical singularity located at $x=-1$, by imposing  
\beq \label{second-constr}
\frac{32 \pi \bar{\D}_\varphi \left[1 - 2 A m + A^2 (p^2 + q^2) \right]}{ ( - 2 + B p )^4 + 2 B^2 q^2 \left[12 + B p (- 4 + B p )\right]} = 2 \pi  \quad .
\eeq

Therefore we remain with a completely regular metric outside the horizon\footnote{Of course the characteristic black hole curvature singularity at $r=0$ remains.}.
This latter regularity constraint, relate the acceleration parameter $A$ with the intensity of the external magnetic field $B$ and the remaining parameters of the black hole conserved charges: the mass $m$, the electric charge $q$ and the magnetic charge $p$, which are though free

\beq \label{A-reg}
      A = \frac{m\left\{ 16 + B^2 (p^2+q^2) [24 + B^2 (p^2+q^2)] \right\}}{8 p B (p^2+q^2) \left[ 4 + B^2 (p^2+q^2) \right] } \pm \sqrt{ \frac{m^2\left\{ 16 + B^2 (p^2+q^2) [24 + B^2 (p^2+q^2)] \right\}^2}{ \left\{8 p B (p^2+q^2) \left[ 4 + B^2 (p^2+q^2) \right] \right\}^2} - 1 } \quad .
\eeq  

From a physical point of view the regularisation of the metric (\ref{elctr-ernst-g})-(\ref{om}) obtained by the constraint (\ref{A-reg}) is interpreted as the removal of the string from the accelerating spacetime. In spite the black hole acceleration is provided by interaction between the external electromagnetic field and the black hole electromagnetic charges. Note that to remove both the singularities from the C-metric, the interaction between the external electromagnetic field and the black hole charge have to be of the same kind.\footnote{For instance,  as it can be seen from (\ref{second-constr}), in the magnetic background embedding considered here, it not possible to remove non-trivially the nodal singularity when $p=0$, but is is possible for $q=0$. When $p=0$ the regularity request leaves only trivial solutions, i.e. $A=0$ or $m=0$ which correspond to cases where naturally there are no axial angular defects: the Reissner-Nordstrom black hole in a Melvin universe or an accelerating Melvin Universe without black hole, respectively.}     \\  
Of course the electromagnetic charges of the black hole are affected by the acceleration and magnetic embeddings, therefore $q$ and $p$ represents the black hole electric and magnetic charges only in the simultaneous  limit of null acceleration and external magnetic field (that is $ A\rightarrow0 $ , $ B \rightarrow 0 $). In fact the actual electric charge can be computed, by a surface integral, as done in the unmagnetised case of section \ref{acc-rev}

\beq \label{QB}
       \mathcal{Q}_B  =   \frac{q \ \left[ 4- B^2 (p^2+q^2) \right] \left[ 16 + 24 B^2 (p^2+q^2) + B^4 (p^2+q^2)^2 \right] }{4 \left[ 1 + 2 A m + A^2 (p^2+q^2) \right] \left[ 16 -32 B p + 24 B^2 (p^2+q^2) -8 B^3 p (p^2+q^2) +B^4 (p^2+q^2)^2  \right] }   \ \ ,
\eeq

while the magnetic monopole charge is

\beq \label{PB}
       \mathcal{P}_B =  \frac{ p \ \left[ 4- B^2 (p^2+q^2) \right]  }{4 \left[1 + 2 A m + A^2 (p^2+q^2)  \right] \left[ 16 -32 B p + 24 B^2 (p^2+q^2) -8 B^3 p (p^2+q^2) +B^4 (p^2+q^2)^2  \right] }   \ \ .
\eeq

The limits for null acceleration or null magnetic field recover the known results of section \ref{acc-rev} and \cite{aliev-gal-cariche}.\\
%
The event horizon area is given by

\beq \label{mag-area}
           \mathcal{A} = \int_0^{2\pi} d\tilde{\varphi} \int_{-1}^1 dx \sqrt{g_{\tilde{\varphi}\tilde{\varphi}} g_{xx}} = \ 4 \pi \D_\varphi \ \frac{r_+^2}{1-A^2 r_+^2} \quad .
\eeq

Note that the dependence of the back hole area from the external magnetic field is implicit, and it only enters in the factor that regulate the period azimuthal angle $\D_\varphi$. When considering regular black holes, $B$ also enters in the value of A according to the constraint (\ref{A-reg}).  \\
In order to take the near horizon limit it will be necessary to know the value of the angular velocity on the event horizon

\beq
         \Omega_J := -\frac{g_{\tilde{t} \tilde{\varphi}}}{g_{\tilde{\varphi}\tilde{\varphi}}} \ \Bigg|_{\tilde{r}=r_+} = - \frac{q B (4 + B r_+ r_-)}{2 \bar{\D}_\varphi r_+} \quad ,
\eeq

and of the Coulomb potential  

\beq
              \Phi_e := - \chi^\m A_\m\bigg|_{\tilde{r}=r_+}  =  \frac{q (4 + B^2 r_+ r_-)}{4 r_+}   \quad .
\eeq

Following exactly the same procedure of section \ref{acc-NH} to obtain the near-horizon geometry for this regularised C-metric we have to pass to the co-rotating frame through the dimensionless coordinate (\ref{change-coordinate}), shift the electric potential as in (\ref{elect-shift}) and perform the limit $\l \rightarrow 0$. As in section \ref{acc-NH}, we are here considering only the extremal configuration. The final near horizon geometry for regular accelerating extremal black hole falls again in the twisted and wrapped product of $AdS_2\times S^2$ class. It can be therefore modelled by the usual near horizon metric (\ref{metric-near}) and electromagnetic one-form \ref{near-A}, where the structure functions are given by

\bea 
           \G(x) &=& \frac{\left[ 4 + B^2 (p^2+q^2) + 4 B p x \right]^2 + (4 B q x)^2}{16 \left[ 1 - A^2 (p^2 + q^2) \right] \left( 1 + A x \sqrt{p^2 + q^2} \right)^2 } \ (p^2+q^2) \quad ,  \label{GBA} \\
           \g(x) &=& \frac{(p^2+q^2) \sqrt{1-x^2} \ \bar{\D}_\varphi^{ext}}{\G(x) \ \sqrt{1-A^2(p^2+q^2)} \left( 1 + A x \sqrt{p^2 + q^2} \right)} \quad ,  \\ 
\a(x) &=& \frac{\sqrt{1-A^2(p^2+q^2)}}{1+xA\sqrt{p^2+q^2}} \qquad \hspace{0.4cm}, \hspace{0.9cm} \k = - \frac{B q \left[ 4 + B^2 (p^2+q^2) \right] r_0^2}{2 (p^2+q^2) \bar{\D}_\varphi^{ext} } \quad ,  \\
e &=& q r_0^2 \ \frac{4 + 3 B^2 (p^2+q^2)}{4 (p^2 + q^2)}   \qquad , \qquad r_0 = \pm \frac{\sqrt{p^2+q^2}}{\sqrt{1-A^2 (p^2+q^2)}} \quad ,  \\
\ell(x) &=& \frac{\left[ - 4 + B^2 (p^2+q^2) \right] \left\{ \left[ 4 + B^2 (p^2+q^2) \right] - \left(4 B x \sqrt{p^2+q^2} \right)^2 \right\} }{ 2 B \left[ 4 + B^2 (p^2+q^2) \right]\left\{\left[ 4 + B^2 (p^2+q^2) + 4 B p x \right]^2 + (4 B q x)^2 \right\}} \  \bar{\D}_\varphi^{ext}  \quad .  \label{GBA-end}
\eea

When the external magnetic field vanishes, $B=0$, eqs. (\ref{GBA})-(\ref{GBA-end}) coincide with (\ref{metric-near})-(\ref{near-ell})\footnote{Remember that in this section we are considering for simplicity $a=0$.}, as expected. Also in this magnetised case the near horizon extreme geometry is different with respect to the Kerr-Newman one, basically because the presence of a non-null acceleration parameter $A$, which generate a conical singularity. Nevertheless imposing the regularity constraints (\ref{reg-const}), (\ref{A-reg}) it is possible to reduce the near horizon extremal geometry of the accelerating magnetised Reissner-Nordstrom black hole to the standard Kerr-Newman extremal near horizon geometry  (\ref{fields01})-(\ref{fields03}). The explicit  isomorphism can be obtained starting with the KN fields  $\G_0 , \g_0 , \a_0, \k_0 $, operate a  the coordinate transformation (\ref{change-coord}) and a ensuing redefinition of the physical parameters of KN the solution as follows

\bea
      a & \mapsto & \mathtt{a}_B \quad \ \ \ , \nn \\
      q &  \mapsto &  \mathcal{Q}^{ext}_B \quad , \\
      p &   \mapsto & \mathcal{P}^{ext}_B \quad , \nn
\eea

where  $\mathcal{Q}^{ext}_B$ and $\mathcal{P}^{ext}_B$ are the extremal electric and magnetic charges (\ref{QB})-(\ref{PB}) 

\beq
 \mathtt{a}_B= \sqrt{\frac{(p^2+q^2)^{3/2}\left\{ 16 B^2 \sqrt{p^2+q^2} - 8 A B p   \left[ 4 + B^2 (p^2+q^2)  \right] + A^2 \sqrt{p^2+q^2} \left[ 4 + B^2 (p^2+q^2)  \right]^2 \ \right\}}{16 \left[ 1- A^2 (p^2+q^2) \right]^3 }} \ . 
\eeq

Exactly as it occurs for the magnetised, non accelerating case \cite{magn-kerr-cft}, the physical parameters change according to the conserved quantities of the solutions. \\
It is natural to take advantage of this map and the extremality condition to infer the value of the mass for this magnetised and accelerating regular extremal black hole 

\beq
       \mathcal{M}^{ext}_B = \sqrt{\mathtt{a}^2_B+(\mathcal{Q}^{ext}_B)^2+(\mathcal{P}^{ext}_B)^2} \quad .
\eeq
In the pure magnetised case \cite{magn-kerr-cft} this insight was useful to discern the mass also away from extremality, as confirmed in \cite{mass-mkn}. It might be of some help also this more general case.

Now we have all the ingredients to compute, according to eq (\ref{cc}), the central charge for the near horizon geometry of the extremal accelerating Reissner-Nordstrom black hole embedded in an external magnetic field
\beq \label{reg-cc}
        c_J = \frac{6 \ \k \ (p^2+q^2)}{1- A^2 (p^2+q^2)} \  \bar{\D}_{\varphi}^{ext} \quad .
\eeq

On the other hand the Frolov-Thorne temperature (\ref{Tphi}) explicitly depends on the intensity of the external magnetic field $B$, since it takes into account the rotational degrees of freedom which comes from the Lorentz interaction between the intrinsic charges of the black hole and the external magnetic field
\beq \label{reg-Tphi}
          T_\varphi = - \frac{1 -A^2 (p^2+q^2)}{B q \pi \left[ 4 + B^2 (p^2+q^2) \right]} \  \bar{\D}_{\varphi}^{ext}   \quad .
\eeq

Finally, as in section \ref{micro}, we can use the Cardy formula (\ref{cardy0}), (\ref{cardy-can-ens}), to compute the entropy of the conformal field theory model dual to the black hole near horizon geometry 
\beq
          \mathcal{S}_{CFT} \ = \  \pi  \bar{\D}_\varphi^{ext} \ \frac{p^2+q^2}{1-A^2 (p^2+q^2)} \  = \ \frac{\mathcal{A}^{ext}}{4} \quad ,     
\eeq 

where the Frolov-Thorne temperature (\ref{reg-Tphi}) was used, as left temperature $T_L$, while as left central charge $c_L$ we referred to eq. (\ref{reg-cc}).\\ 
Remarkably the CFT entropy coincide with a quarter of the extremal black hole area $\mathcal{A}^{ext}$, that is the extremal limit of eq (\ref{mag-area}). Hence the entropy of the dual two dimensional conformal field model corresponds to the standard Bekestein-Hawking black hole entropy.\\
Therefore, also in this regular case, where the black hole is non-trivially deformed by the presence of acceleration and of an external magnetic field, the Kerr/CFT correspondence has shown to hold at extremality. \\
The second dual conformal picture, where the electromagnetic $U(1)$ gauge symmetry is exploited, can be pursued also in presence of the external magnetic field as in the section \ref{micro}. The resulting central charge and Frolov-Thorne temperature are respectively read

\bea
         c_Q      &=&  \frac{3 q (p^2+q^2)\left[ 4 + 3 B^2 (p^2+q^2) \right]}{2 R_\psi \left[1-A^2(p^2+q^2) \right]^2 } \bar{\D}_\varphi     \\
         T_\psi   &=&   \frac{2 R_\psi \left[1-A^2(p^2+q^2) \right] }{\pi q \left[ 4 + 3 B^2 (p^2+q^2) \right] }     
\eea

Therefore thanks to the Cardy formula (\ref{cardy-can-ens}) we can confirm that the gravitational entropy can be reproduced also in this alternative dual picture, as in (\ref{entro-alt}).\\
It is worth to notice that, as already observed in \cite{magn-kerr-cft}, the presence of the external electromagnetic field improves the applicability of the Cardy formula, in booth the conformal pictures. That happens because, through the factor $\bar{\D}_\psi$, for a specific range of parameters, it is possible to fulfil the sufficient condition for the applicability of the Cardy formula, which in having the temperature much larger with respect to the central charge. This means that there are a large number of excited degrees of freedom.  \\
While presence of the regularising electromagnetic background makes the near horizon extremal geometry isomorphic to the extremal Kerr-Newman one, the fact that the Kerr/CFT works still remains not trivial because some quantities, such as the left temperature $T_\varphi$ of the dual conformal field model, are computed outside the near horizon limit.\\
Moreover it can be shown that the Kerr/CFT formalism works well also for more complicated generalisation of these accelerating regularised black holes, such as the one with not null $a=0$\footnote{Actually, by an electromagnetic duality transformation, everything can be further generalised in presence of external electric field, as well.}. Since there are not any additional conceptual issues, with respect to the example presented in this section, we will avoid to discuss it here.\\
On the other hand, the addition of the cosmological constant, in the regularised case is not as easy as in the accelerating but unmagnetised case because a magnetising Harrison transformation in presence of the cosmological constant is not known at the moment \cite{marcoa-lambda}.\\

The study of the non-extremal case in this magnetised and accelerating scenario would be very interesting, but it is not clear if the standard methods based on the separability of a non-interacting probe scalar field can be applied. The main problem is that it is knot known if its scalar wave equation can be decoupled in a radial and angular part and therefore if the hidden $SL(2,\mathbb{R}) \times SL(2,\mathbb{R})$ symmetry can be exploited. But also the usual assumptions on the value of the central charge would reveal to be problematic because, as we have seen in (\ref{costr-D}) the constraint on the period of the azimuthal angle is not compatible with the regularity constraint (\ref{reg-const}). Therefore an independent computation of the central charges for accelerating black hole become even more necessary than in the Kerr-Newman case.\\

\section{Comments and Conclusions}

In this paper we analysed the near horizon geometry of accelerating Kerr-Newman black holes. We have verified that at extremality the near horizon geometry can be written as a warped and twisted product of $AdS_2 \times S^2$, but it is different from the extremal Kerr-Newman  near horizon geometry. Thus the presence of the acceleration modifies the near horizon region, unlike  what occurs with other deformations of the Kerr-Newman spacetime, such as the external magnetic field. This extremal near horizon geometry possesses the $SL(2,\mathbb{R})$ symmetry, which can be exploited by the Kerr/CFT correspondence. Indeed, at extremality, all the methods of the Kerr/CFT can be smoothly applied in presence of the acceleration. We found how the acceleration enters in the central charge of the asymptotic near horizon geometry and how it deforms the Frolov-Thorne temperature. Thus it was possible, according to the Kerr/CFT prescription, to map the gravitational system into a two-dimensional conformal field theory model. We confirmed that the entropy, computed with the tools provided by the CFT, matches the gravitational Bekenstein-Hawking temperature for the accelerating and rotating extremal black hole.\\
We have explicitly shown how these results hold both for standard rotating C-metrics, which present conical singularity, and for regularised rotating and accelerating Reissner-Nordstrom black holes. Actually the presence of the regularising external magnetic field improves the correspondence with the conformal field theory model, enhancing  the applicability of the Cardy formula.\\
The extremal regularised metric is isomorphic, in the near horizon limit, to the extremal Kerr-Newman one, the map is explicitly presented. Hence the deviation from the Kerr-Newman metric is not due to the acceleration, but to the source of the acceleration. The cosmic string driving the acceleration in the first case generates a conical singularity causing the different near horizon behaviour.   \\
Note that many of the difficulties characteristic of these magnetised and accelerating spacetimes, typically related with the non-constant curvature asymptotic, were avoided just because we were mainly dealing with near horizon quantities. This fact remarks once more the fundamental role played by the event horizon in the physics of black holes.\\
Further generalisations, such as the inclusion of the cosmological constant to the accelerating picture, are trivial at least at extremality. What is less trivial, in presence of the black hole deformations considered in this paper, is the non-extremal picture. Indeed some fundamental symmetries based on the separability of the wave equation of a probe scalar field on these accelerating black hole backgrounds are preserved. However it is not clear how to implement some of the ad-hoc assumptions on the nature of the central charges, typical of the non-extremal limit. Neither it is clear how to compute the central charges away from extremality, but this is a known issue in the formulation the Kerr/CFT correspondence, which is independent from the presence of the acceleration or external electromagnetic fields. \\
Finally we remark that we were able to provide, for the first time in the literature a value for the mass of accelerating black holes that fulfil the standard first law of black hole thermodynamics, without extra assumptions. Would be interesting to confirm this result by direct computation with methods that do not assume the validity of the first laws. The canonical frame provided in section \ref{mass-acc} could be useful in that direction. This will also clarify the uniqueness of the proposed mass.  \\

\section*{Acknowledgements}
\small I would like to thank Geoffrey Compere, James Lucietti, Roberto Oliveri for fruitful discussions and Fabrizio Canfora for computational support.
\small This work has been funded by the Conicyt - PAI grant 79150061.\\
\normalsize

%

\end{document}